\documentclass[aps,prb,twocolumn,superscriptaddress,nolinenumbers,nobibnotes,nodoi,noeprint]{revtex4-2}

\usepackage{amsmath,amssymb,physics,bm,siunitx}
\usepackage{xcolor,graphicx}
\usepackage{soul}

\usepackage[papersize={8.5in,11in}]{geometry}
\usepackage[colorlinks=true]{hyperref}
\hypersetup{
    bookmarks=true,         
    unicode=false,          
    pdftoolbar=true,        
    pdfmenubar=true,        
    pdffitwindow=false,     
    pdfstartview={FitH},    
    pdfkeywords={keyword1} {key2} {key3}, 
    pdfnewwindow=true,      
    colorlinks=true,       
    linkcolor=magenta, 
    citecolor=blue,        
    filecolor=magenta,      
    urlcolor=blue           
} 

\def \nn{\nonumber \\}

\AtBeginDocument{\RenewCommandCopy\qty\SI}
\begin{document}

\title{Hong--Ou--Mandel interferometry for fractional excitations:\\ Unified framework and dip width scaling}
\author{Aleksander Latyshev$^1$, Imen Taktak$^2$, Ipsita Mandal$^3$ and In\`es Safi}
 \affiliation{$1$: Laboratoire de Physique des Solides (UMR 5802), CNRS-Universit\'e Paris-Saclay,
B\^atiment 510, 91405 Orsay, France\\
$2$: SPEC-CEA Orme des Merisiers, 91190 Gif sur Yvette\\
$3$: Department of Physics, Shiv Nadar Institution of Eminence (SNIoE), Gautam Buddha Nagar, Uttar Pradesh 201314, India}

\begin{abstract}
Extending Hong--Ou--Mandel (HOM) interferometry  to the fractional quantum Hall effect (FQHE) promises direct access to anyonic statistics, yet remains challenging: on-demand anyon injection is hindered by integer-charged minimal excitations, and recent HOM experiments in the FQHE lack a fully consistent theoretical framework.
Here we provide a general theory of time-resolved HOM interferometry in quantum Hall systems. Combining the nonequilibrium bosonized edge theory (NEBET) with the unifying non-equilibrium perturbative theory (UNEPT), we derive exact and perturbative relations obeyed by the relevant cross-correlations of  chiral currents valid for spatially extended tunneling operators and generic quadratic edge dynamics. Then, within the Tomonaga–Luttinger liquid framework, we analyze the width of the HOM dip for injected pulses carrying integer and fractional charges. We show that it is governed by the width of the pulses and, for the fractional charge, by a non-trivial power-law behavior of the scaling dimension $\delta$. Our results establish a robust theoretical foundation for interpreting recent experiments on anyonic statistics and electronic interferometry in the quantum Hall regime.

\end{abstract}

\maketitle

\emph{Introduction.---}
Manipulating individual quasiparticles in the quantum Hall regime has opened a pathway to electronic quantum optics (EQO) \cite{ines_epj,feve_07_on_demand,klich_levitov,glattli_levitons_nature_13,quantum_optics_rev_grenier_2011}, where single electrons emitted in ballistic chiral edges play the role of single photons in optical media. Fermi statistics and strong interactions endow EQO with features absent in photon optics. A central tool in this field is interferometry, in particular Hong--Ou--Mandel (HOM) geometries where synchronized sources emit excitations that collide at a quantum point contact (QPC). HOM setups have provided evidence for charge fractionalization \cite{ines_epj,fractionnalisation_eugene_mach_zhender_prb_2008,plasmon_ines_IQHE_Hong-Ou-Mandel_feve_Nature_2015_cite,plasmon_artificial_TLL_experiment_fujisawa_nature_2014_cite,sassetti_levitons_IQHE_2020} and enabled electronic-state tomography \cite{tomography_Grenier_2011,tomography_glattli_2014,tomography_degiovanni_feve_2019}. While highly successful in the integer quantum Hall regime, where deterministic single-electron sources are well established \cite{feve_07_on_demand,klich_levitov, glattli_levitons_nature_13}, HOM interferometry remains far less developed in the fractional quantum Hall effect (FQHE). While the determination of the fractional charge, often based on DC shot noise \cite{kane_fisher_noise,ines_resonance,saminad,heiblum_frac_97}, has more recently benefited from robust time-dependent transport methods \cite{ines_eugene,ines_degiovanni_2016,ines_PRB_R_noise_2020,ines_PRB_2019,glattli_photo,ines_gwendal,ines_photo_noise_PRB_2022,Feldman_review_2021}, probing fractional statistics still relies primarily on \textsc{dc} transport \cite{ines_prl,kim_hbt,fractional_statistics_theory_2016,fractional_statistics_gwendal_science_2020,pierre_anyons_PRX_2023,fractional_statistics_gwendal_PRX_2023,fractional_statistics_anyons_FQHE_heiblum_MZI_2_5_Nature_physics_2023,manfra_braiding_anyons_2022,manfra_2_5_braiding_FPI_FQHE_PRX_2023}. Attempts at on-demand anyon injection have so far been hindered by the fact that driven quantum dots emit only electrons and that minimal excitations generated by Lorentzian voltage pulses\cite{klich_levitov}  necessarily carry integer charge \cite{glattli_levitons_physica_2017}.

Well before the advent of electron quantum optics in 2007 \cite{feve_07_on_demand}, one of the present authors anticipated the possibility of shaping propagating plasmonic waves with charge set by time-dependent voltages \cite{ines_schulz,ines_ann,ines_epj}. This non-equilibrium bosonization framework related finite-frequency admittance to the plasmon scattering matrix, providing concepts and tools that underlie modern time-resolved experiments in edge states \cite{plasmon_artificial_TLL_experiment_fujisawa_nature_2014_cite,plasmon_ines_experiment_fujisawa_PRB_2014_cite,feve_degiovanni_ines_bartolomei_FPI_2025,degiovanni_radar_ines_ANR_MZI_PRX_2025}. This line of thought culminated in a recent experiment \cite{feve_anyons_martin_Science_2025}, where an HOM setup with injected fractional pulses was used to extract both the scaling dimension $\delta$ and the braiding phase $\theta$ of anyons. The theoretical analysis supporting this work \cite{martin_anyons_dip_HOM_PRL_2023} is based on a Tomonaga--Luttinger liquid (TLL) model \cite{FQHE_review_wen_1992}. While TLL predictions are consistent with a recent experiment \cite{manfra_TLL_graphene_2025}, however, they disagree with experiments determining the fractional charge \cite{glattli_photo_2018,ines_gwendal} and statistics \cite{fractional_statistics_gwendal_science_2020,pierre_anyons_PRX_2023,fractional_statistics_gwendal_PRX_2023,kyrylo_note}. Two complementary frameworks address time-dependent transport beyond the TLL paradigm: the nonequilibrium bosonized edge theory (NEBET), valid for arbitrary ranges and forms of interactions \cite{ines_bena,ines_bena_crepieux,trauzettel_04,gu_unpublished}, and the unifying nonequilibrium perturbative theory (UNEPT) \cite{ines_eugene,ines_PRB_2019,ines_photo_noise_PRB_2022}, which applies to broad classes of correlated systems including quantum Hall edges, Josephson junctions, and their dual phase-slip junctions. UNEPT, perturbative in the time-dependent Hamiltonian term, yields universal fluctuation-dissipation relations linking photo-assisted current and noise to their \textsc{dc} counterparts. It has been successfully used to analyze HOM-type experiments for injected electrons in a unified manner between integer and FQHE \cite{glattli_imen_2022}.

In this Letter, we revisit HOM interferometry in the FQHE by combining these two theories. We make two contributions. First, we derive two general relations—one exact, the other perturbative—for cross-correlations of outgoing chiral currents, based on a general analysis of photo-assisted noise \cite{exact_FDR_unpublished}. This goes beyond and revisits all previous HOM works \cite{martin_sassetti_prl_2017,martin_sassetti_crystallization_TD_PRB_2018,martin_sassetti_Hong-Ou-Mandel_multiple_levitons_EPJ_2018,martin_sassetti_Hong-Ou-Mandel_heat_PRB_2018,martin_fractional_statistics_dip_HOM_PRL_2023,Sassetti_Hong-Ou-Mandel}, which focused solely on backscattering noise and assumed local tunneling within the TLL model. Second, we carefully analyze the short-pulse protocols implemented in Refs.~\cite{martin_anyons_dip_HOM_PRL_2023,fractional_statistics_gwendal_science_2020} under local backscattering with dimension $\delta$, showing that their interpretation requires a refined theoretical framework, which also addresses the validity domain of our analysis~\cite{ines_imen_2025}. Our results establish a comprehensive, drive-agnostic description of photo-assisted HOM interferometry in the FQHE, and provide concrete predictions for the width and behavior of the HOM dip—resolving discrepancies and offering robust signatures for time-resolved anyon experiments.

\begin{figure}[t!]
\centering
\includegraphics[width= 0.99 \linewidth]{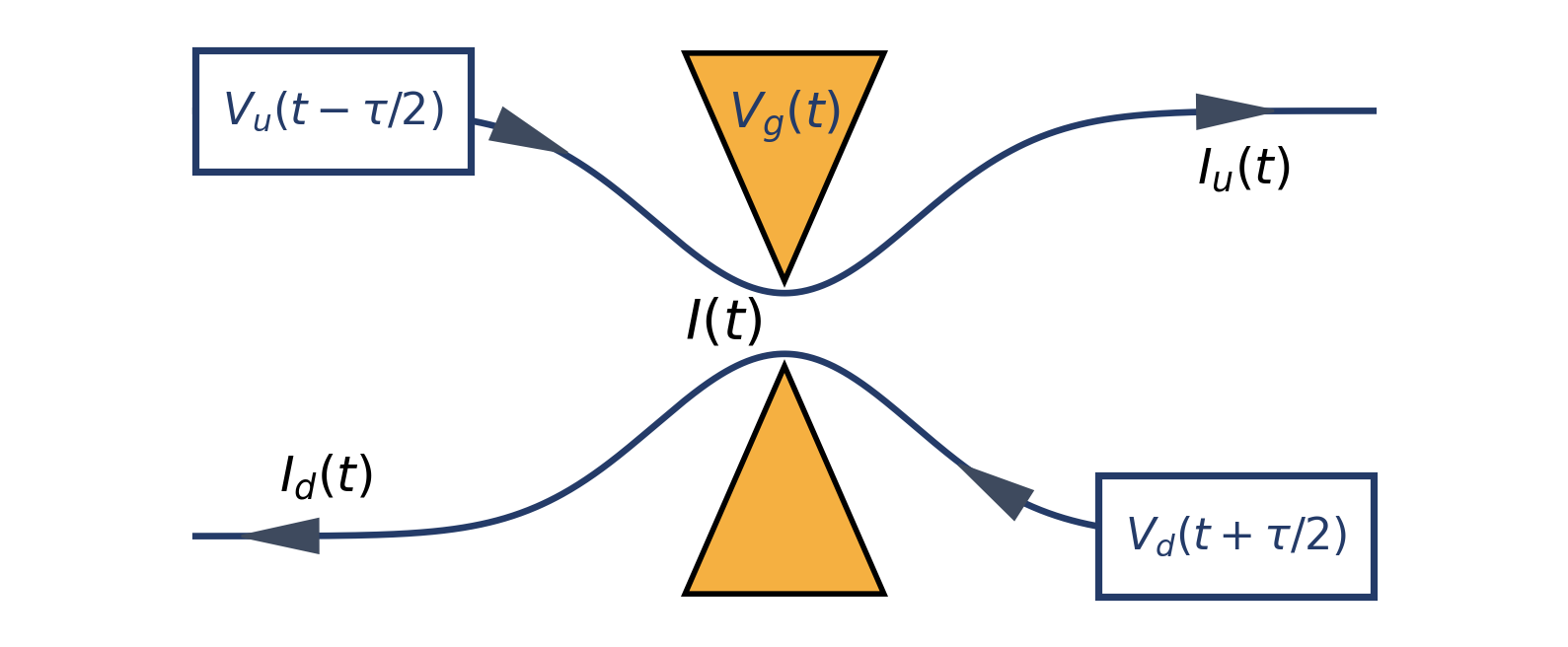}
\caption{A QPC in the quantum Hall regime at an integer or fractional filling factor $\nu$. We focus here on {edges such that each harbors only a single chiral mode. While $V_u$ and $V_d$ denote the reservoir voltages, $V_g$ denotes a gate voltage.} Both the reservoir and gate voltages can be time-dependent. $I_u(t)$ and $I_d(t)$ denote the outgoing chiral current operators in the upper and lower edges [cf. Eq.~\eqref{eq:chiral_current}], respectively, and $I(t)$ represents the backscattering-current operator in Eq.~\eqref{eq:backscattering_current}.
\label{fig_cross}}
\end{figure}

\emph{Model and exact NEBET relations.---}
In this Letter, we consider a single chiral edge mode of a fractional quantum Hall state at filling factor $\nu$, described by bosonic fields $\phi_{u,d}$, where $u$ and $d$ denote the upper and lower edges of the Hall bar, respectively. The edge dynamics are governed by a quadratic bosonized Hamiltonian $\mathcal{H}_0$ in terms of $\phi_{u,d}$, without assuming the specific form of a (chiral) Tomonaga–Luttinger liquid. Quasiparticle backscattering is represented by a generic, possibly spatially extended operator $A$, driven by a complex time-dependent function $\tilde{p}(t,\tau)$. We allow here for an explicit dependence on a time shift $\tau$, which in the HOM setup corresponds to the relative delay between two sources. All time-dependent forces are incorporated—either through a Keldysh gauge transformation or, equivalently, by evolving time-dependent boundary conditions \cite{ines_epj} --- into $\tilde{p}(t,\tau)$. Importantly, the magnitude $|\tilde{p}(\tau,t)|$ is not constrained to unity, which permits amplitude modulation, for instance from a time-dependent gate voltage (see Fig.~\ref{fig_cross}).

Accordingly, the full time-dependent Hamiltonian takes the form
\begin{align}
\label{eq:hamiltonian}
\mathcal{H}(t)=\mathcal{H}_0+\mathrm{e}^{-i\omega_J t}\,\tilde{p}(t,\tau)\,A+\mathrm{e}^{i\omega_J t}\,\tilde{p}^{*}(t,\tau)\,A^{\dagger},
\end{align}
where $\omega_J$ is a Josephson-like frequency. Although not required in full generality, $\omega_J$ often satisfies the Josephson-type relation $\omega_J = e^* V/\hbar$, where $V$ denotes the applied dc voltage drop between the upper and lower edges and $e^*$ the transferred quasiparticle charge.

Two current operators will play a central role. First, the quasiparticle tunneling current at the QPC,
\begin{align}
\label{eq:backscattering_current}
I(t) = -\,i\,\frac{e^{*}}{\hbar}\!\left[\mathrm{e}^{-i\omega_J t}\,\tilde{p}(t,\tau)\,A-\mathrm{e}^{i\omega_J t}\,\tilde{p}^{*}(t,\tau)\,A^{\dagger}\right],
\end{align}
whose fluctuations define the photo-assisted tunneling (backscattering) noise,
\begin{align}\label{eq:SB_ph_definition}
 {S}_{ph}(\omega_{J},\tau)=\int_{-\infty}^{\infty}\!\!dt \!\!\int_{-\infty}^{\infty}\!\!\!ds\,  
 \left< \delta I_{\mathcal{H}}\!\left(t-\tfrac{s}{2}\right)\,
 \delta I_{\mathcal{H}}\!\left(t+\tfrac{s}{2}\right)\right>,
\end{align} 
where $\delta I_{\mathcal{H}}(t)=I_{\mathcal{H}}(t)-\langle I_{\mathcal{H}}(t)\rangle$. Here the subscript $\mathcal{H}$ indicates the Heisenberg representation, and the average is taken over a thermal initial state $\rho_{th}\propto e^{-\beta\mathcal{H}_0}$ with electronic temperature $T=1/\beta$. 
Second, we will consider the experimentally accessible chiral edge currents, 
\begin{align}
\label{eq:chiral_current}
I_{u,d}(x,t)=v\,\partial_x \phi_{u,d}(x,t)\,,
\end{align}
where $v$ is the edge magnetoplasmon velocity. The corresponding cross-correlations are defined as
\begin{align}
\label{eq:Schiral_ph_definition}
 & {S}^{ud}_{ph}(\omega_{J},\tau) 
 \nn & = 
 \int_{-\infty}^{\infty}\!\!dt \!\!\int_{-\infty}^{\infty}\!\!\!ds\,
 \left< \delta I_{u,\mathcal{H}}\!\left(x_u,t-\tfrac{s}{2}\right)\,
 \delta I_{d,\mathcal{H}}\!\left(x_d,t+\tfrac{s}{2}\right)\right>,
\end{align} 
with $\delta I_{\zeta,\mathcal{H}}(x_{\zeta},t)= I_{\zeta,\mathcal{H}}(x_{\zeta},t)-\langle I_{\zeta,\mathcal{H}}(x_{\zeta},t)\rangle$ and $\zeta=u,d$ and $x_{u,d}$ are the upper and lower edge measurement points.
We also introduce the photo-assisted tunneling current,
\begin{align}
\label{eq:I_B_ph_definition}
 {I}_{ph}(\omega_{J},\tau) =
\int_{-\infty}^{\infty}\!\!dt \,\langle I_{\mathcal{H}}(t)\rangle,
\end{align} 
with an analogous definition for the chiral-current averages. {With this definition, Kirchhoff’s laws are satisfied---} for instance,
\begin{align}
 {I}_{u,ph}(\omega_{J},\tau)=
 e\,\nu \,\omega_J - {I}_{ph}(\omega_{J},\tau)\,.
\end{align}
We further define the backscattering photoconductance,
\begin{align}
G_{ph}(\omega_J,\tau)=\frac{\partial I_{ph}(\omega_J,\tau)}{\partial \omega_J},
\end{align}
which has the dimension of an electric charge. The temperature-dependence is implicit in all these observables.  

Extending the NEBET developed in Refs.~\cite{ines_bena,ines_bena_crepieux,gu_unpublished} for constant voltages to time-dependent drives, we obtain an exact relation connecting photo-assisted cross-correlations of chiral currents to the backscattering photo-assisted noise \cite{exact_FDR_unpublished}:
\begin{align}
\label{eq:exact_cross_photo}
 S_{ph}^{ud}(\omega_J,\tau)=S_{ph}(\omega_J,\tau)-2\,\omega_{th}\,G_{ph}(\omega_J,\tau),
\end{align}
where we introduce the thermal frequency $$\omega_{th}=\frac{k_B T}{\hbar}.$$  This is a central relation of this Letter, which is, to our knowledge, the first exact expression for cross-correlations under time-dependent drives \footnote{In the zero-frequency limit, Eq.~\eqref{eq:exact_cross_dc} coincides with the modified excess noise defined in Ref.~\cite{ines_philippe_group, *safi09}, which differs from the standard excess noise: $\Delta S_{dc}(\omega_J) = S_{dc}(\omega_J) - S_{dc}(0)$,
with $S_{dc}(0) = 2 \,\omega_T\, G_{dc}(0)$ (FDT). Both coincide only for a linear QPC, where $G_{dc}(\omega_J) = G_{dc}(0)$, in which case $S^{ud}_{dc}(\omega_J, T) = \Delta S_{dc}(\omega_J, T)$.}. 
Auto-correlations take a similar form, with an additional contribution from equilibrium noise in the absence of the QPC. Importantly, this result extends beyond Laughlin states: it requires only that $\mathcal{H}_0$ be quadratic in bosonic fields (allowing for arbitrary inter-edge interactions) and that $A$ represent a generic, possibly spatially extended, tunneling operator.  

Within the same framework, NEBET—originally formulated at finite frequency and dc voltages \cite{ines_bena_crepieux,ines_bena,dolcini_07}—provides exact relations between cross-correlations and backscattering noise. When combined with generalized linear response theory \cite{ines_philippe}, this yields an exact expression in the dc regime \cite{gu_unpublished}:
\begin{align}
\label{eq:exact_cross_dc}
S^{ud}_{dc}(\omega_J) =
S_{dc}(\omega_J) - 2 \,\omega_{th}\, G_{dc}(\omega_J)\,,
\end{align}
where the dc differential conductance is defined as $G_{dc}(\omega_J)={\partial I_{dc}(\omega_J)}/{\partial \omega_J}$.  The validity of bosonisation is restricted to energies below a UV cutoff $\omega_c$, thus $\omega_{th} \ll  \omega_c$. The dc noise and current are the stationary analogs of Eqs.~\eqref{eq:SB_ph_definition}, \eqref{eq:Schiral_ph_definition}, and \eqref{eq:I_B_ph_definition}, corresponding to $\tilde{p}(\tau,t)=1$. Notice that for a linear dc current, $S^{ud}_{dc}(\omega_J)$ reduces to the excess noise.  At equilibrium ($\omega_J=0$), the fluctuation–dissipation theorem (FDT) gives
\begin{align}
\label{eq:exact_cross_dc_equilibrium}
S^{ud}_{dc}(0) = 0 \,.
\end{align}
This cancellation is, however, specific to an initial thermal distribution; being tied to the equilibrium FDT, one can have nonvanishing cross correlations for nonequilibrium distributions (such as in the anyon collider, see Ref.~\cite{ines_PRB_R_noise_2020}). 

The relation in Eq.~\eqref{eq:exact_cross_photo} applies to arbitrary drive profiles, which may differ between right and left sources. Let us now restrict it 
to the specific HOM geometry of Fig.~\ref{fig_cross}, with two time-dependent sources operating with a time delay $\tau$. In this case, the profile in Eq.~\eqref{eq:hamiltonian} takes the form of
\begin{align}
\label{eq:pi}
\tilde{p}(t,\tau) =  e^{-i \, \left[\varphi(t-\tau/2)-\varphi(t+\tau/2)\right]}\,,
\end{align}
where $\varphi(t)$ describes the ac phase applied to each source. In line with experimental practice, we assume identical dc voltages in both sources, so that $\omega_J=0$, and we introduce the subscript \emph{HOM} for the corresponding cross-correlated PASN:
\begin{align}
S^{HOM}(\tau)=S^{ud}_{ph}(\omega_J=0,\tau)\,.
\end{align}
\paragraph{The HOM dip: general limitations}
Let us now address the HOM dip that one generally expects, keeping unspecified the shape of $\varphi(t)$ and the form of the bosonized Hamiltonian in Eq.~\eqref{eq:hamiltonian}. From Eq.~\eqref{eq:exact_cross_dc_equilibrium}, one immediately obtains the universal and nonperturbative result that the HOM noise vanishes at zero delay, at any temperature. This is a consequence of gauge invariance, as one is back to the dc regime, as $\tilde{p}(t,\tau=0)=1$:
\begin{align}
\label{eq:exact_cross_ph_equilibrium}
S^{ud}_{ph}(\tau=0) = S^{ud}_{dc}(\omega_J=0) = 0\,.
\end{align}
When the sources inject single electrons, the vanishing of the HOM signal at $\tau=0$ can be interpreted as a signature of antibunching: synchronized electrons cannot collide at the QPC. For Lorentzian-shaped voltages, the HOM signal factorizes into a temperature-dependent prefactor and $1-g(\tau)$, where $g(\tau)$ denotes the overlap of the single-electron wavefunctions~\cite{glattli_levitons_physica_2017}, resulting in a temperature-independent width. Within the UNEPT, we can show (as will be detailed elsewhere) that this property—originally derived for independent electrons—remains robust against both the range and strength of interactions. Thus, in this case, the HOM dip shape is fully determined by that of the Lorentzian pulses.

For injected excitations carrying a fractional charge $q$ and braiding with anyons with statistical phase $\theta$, one may instead expect, in a speculative manner, a prefactor $1+\cos\pi\theta$, which vanishes at $\theta=\pi$ and equals $2$ for bosons, reflecting bunching. The fact that Eq.~\eqref{eq:exact_cross_ph_equilibrium} still predicts a strictly vanishing HOM signal for injected fractional charges mimicking anyons may thus be reconsidered as a consequence of the impossibility of strictly reaching $\tau=0$, as will be further clarified within the UNEPT.  

On the one hand, let us assume extremely sharp pulses,  
\begin{align}\label{eq:V_sharp}
\varphi(t) = 2\pi q \;\theta(t),
\end{align}
each carrying a fractional charge $q$ and separated by a time delay $\tau$. In this case, the validity of the low-energy effective theory below a UV energy cutoff $\omega_c$ imposes the condition $\tau\omega_c \gg 1$, which prevents one from taking $\tau=0$.  

On the other hand, consider a finite-width plasmonic pulse, generated by a single rectangular voltage pulse  of duration $a$, thus a phase:
\begin{align}\label{eq:phi_finite_width}
\varphi(t) = \kappa \big[ t\,\theta(t) - (t-a)\,\theta(t-a) \big],
\end{align}
where $\kappa = 2\pi q / a$. In this case, one can choose $a\omega_c \gg 1$ so that $\tau=0$ remains compatible with the UV cutoff. However, the physical interpretation becomes less clear if the time delay $\tau$ is shorter than the pulse width $a$, as the two pulses strongly overlap and cannot be distinguished.  These limitations in accessing $\tau=0$, which apply equally to single electrons, have, to our knowledge, not been explicitly addressed so far.


\paragraph{{Perturbative relations.---}}
We now focus on the weak backscattering regime, where tunneling of fractional charges dominates, described by a small tunneling operator $A$. In this regime, the relations between quasiparticle tunneling photoconductance and noise, $O=S,G$, can be obtained from the UNEPT, specified to zero dc voltage \cite{ines_eugene, ines_cond_mat, ines_photo_noise_PRB_2022}:
\begin{align}\label{eq:UNEPT_ph}
O_{ph}(\omega_J=0,\tau) = \int_{-\infty}^{+\infty} d\omega,\ |\tilde{p}(\omega,\tau)|^2\, O_{dc}(\omega_J=\omega)\,.
\end{align}
Combining this with Eq.~\eqref{eq:exact_cross_photo}, we establish a formal relation between the HOM noise and its nonequilibrium dc counterpart as follows:
\begin{align}
\label{eq:S_perturbative_HOM}
S^{HOM}(\tau) = \int_{-\infty}^{+\infty} d\omega\, |\tilde{p}(\omega,\tau)|^2\, S_{dc}^{ud}(\omega_J=\omega).
\end{align}
Thus, the sideband transmission picture, already generalized to strongly correlated systems by the UNEPT through Eq.~\eqref{eq:UNEPT_ph}, is now extended to chiral current correlations. The HOM noise is expressed as a continuous superposition of contributions, each containing two factors: one associated with the drive $\tilde{p}(t,\tau)$, and the other given by $S^{ud}_{dc}(\omega_J)$, which retains the signature of the underlying Hamiltonian. All dependence on the time delay $\tau$ enters exclusively through $\tilde{p}(\omega,\tau)$. Remarkably, this perturbative relation can also be extended to initial nonequilibrium distributions, such as those induced by temperature gradients or in the “anyon collider,” as will be shown in a separate work \cite{exact_FDR_unpublished}.

\begin{figure}[t!]
    \centering
\includegraphics[width= 0.99 \linewidth]{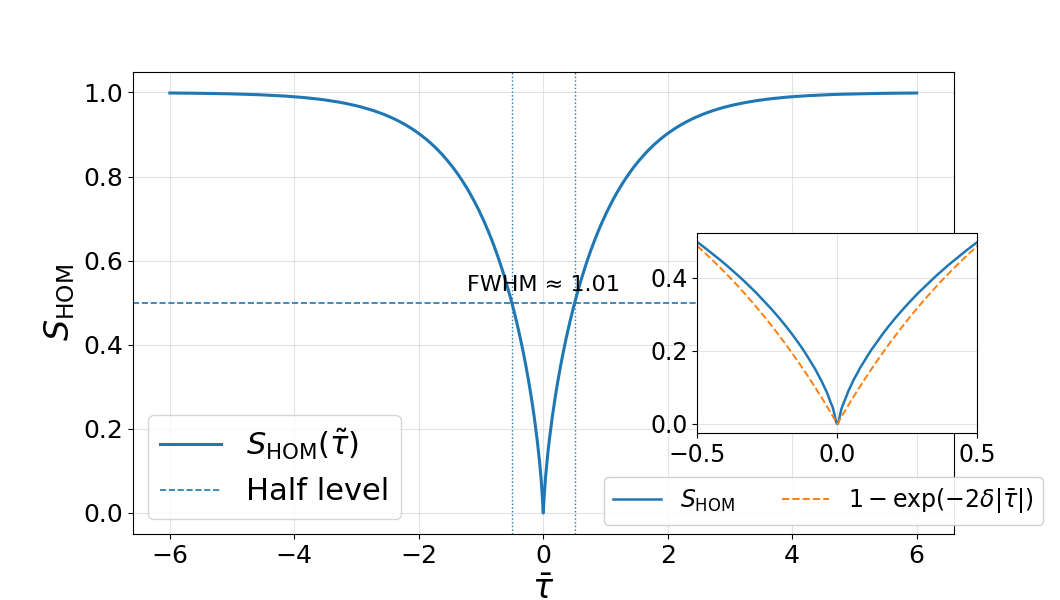}
    \caption{Normalized HOM noise $S^{\mathrm{HOM}}$ as a function of $\bar{\tau}= \pi \omega_{th} \,\tau$ for sharp pulses and for $\delta=2/3$. It is independent of the value of the injected non-integer dimensionless charge $q$. The inset zooms on the vicinity of $\tilde{\tau}=0$ and compares with $\left[ 1-\exp(- 2\delta |\bar{\tau}|) \right ] $ in Ref.~\cite{martin_anyons_dip_HOM_PRL_2023}\label{fig:Fig.2}}
\end{figure}

\paragraph{{Application to two incident sharp pulses.---}}
Following Refs.~\cite{ines_schulz,ines_ann} and the considerations of Ref.~\cite{martin_fractional_statistics_dip_HOM_PRL_2023}, we now consider counter-phased plasmonic pulses and derive the explicit form of the kernel $|\tilde{p}(\omega,\tau)|^2$ entering Eq.~\eqref{eq:S_perturbative_HOM}, without specifying the underlying bosonized Hamiltonian. We first address the case of extremely sharp pulses defined in Eq.~\eqref{eq:V_sharp}. For such a drive one obtains, at nonzero frequency:
\begin{align}\label{eq:psharp}
 \tilde{p}(\omega,\tau)
= \left(e^{2\pi i q}-1\right)
\frac{2\sin\!\left(\omega \tau/2\right)}{\omega}\,.
\end{align}
This leads to a universal oscillatory behavior of all photoassisted observables, Eqs.~\eqref{eq:UNEPT_ph},\eqref{eq:S_perturbative_HOM}, a feature also present in Hanbury Brown–Twiss configurations, as noted in Refs.~\cite{ines_eugene,ines_PRB_2019} for periodic pulses and in Ref.~\cite{lucas_ines_unpublished_2025}  for a single pulse. We normalize Eq.~\eqref{eq:S_perturbative_HOM} by its large-delay value $\tau\!\to\!\infty$, where the pulses do not overlap. In this limit, the sources act independently and the noise reduces to the sum of two Hanbury Brown–Twiss contributions~\cite{martin_fractional_statistics_dip_HOM_PRL_2023}. Remarkably, for noninteger $q$ the normalized HOM signal $S^{\mathrm{HOM}}(\tau)$ is \emph{independent of} $q$, since the prefactor $\lvert e^{2\pi i q}-1\rvert^{2}$ cancels in the ratio. This independence holds irrespective of the Hamiltonian in Eq.~\eqref{eq:exact_cross_dc} or of the possible spatial extension of quasiparticle tunneling. However, the limit $q\to \text{integer}$ is singular: in that case the HOM signal vanishes identically~\cite{lucas_ines_unpublished_2025}.  
These limitations motivate us to go beyond $\delta$-like pulses and consider more realistic plasmonic pulses of finite temporal width $a$, characterized by the phase in Eq.~\eqref{eq:phi_finite_width}. A further motivation, already discussed in relation to the exact NEBET result Eq.~\eqref{eq:exact_cross_ph_equilibrium}, concerns the requirement of bosonization validity at low energies, below the ultraviolet cutoff $\omega_c$. While Eq.~\eqref{eq:S_perturbative_HOM} is general, its use within a bosonized theory requires that the integrand decays sufficiently fast above $\omega_c$ \cite{ines_imen_2025}, unlike the UNEPT relations for backscattering noise and conductance in Eq.~\eqref{eq:UNEPT_ph}, which do not rely on bosonization. Consequently, although HOM noise strictly vanishes at $\tau=0$ by Eq.~\eqref{eq:exact_cross_ph_equilibrium}, this short-delay regime lies outside the range of validity of the bosonized description. A natural way to regularize this regime is to consider pulses of finite width $a$, much larger than the short-time cutoff $2\pi/\omega_c$. Such a single rectangular pulse can also be viewed as the limiting case of the periodic drives employed in Ref.~\cite{feve_anyons_martin_Science_2025}, obtained in the infinite-period limit.

Fourier transforming Eq.~\eqref{eq:pi}, at finite frequency, yields
\begin{align}
& \tilde{p}(\omega,\tau\leq a) \nn
& =
\frac{2 \,e^{\frac{i\,\omega \,a} {2} }} {\omega}
\left [ e^{i \, \kappa \, \tau}
\sin\Bigl(\omega\, \frac{a-\tau}{2}\Bigr)
-\sin\Bigl(\omega \, \frac{a+\tau}{2}\Bigr) \right]
\nn & \quad +
2\, e^{\frac{i\,\kappa\, \tau} {2}}
\left[
\frac{\sin\bigl((\omega+\kappa)\tfrac{\tau}{2}\bigr)}{\omega+\kappa}
+e^{i \, \omega \, a}\,
\frac{\sin\bigl((\omega-\kappa)\tfrac{\tau}{2}\bigr)}{\omega-\kappa}
\right],
\end{align}
with the property $\tilde{p}(\omega,0)=0$. For $\tau>a$, one finds
\begin{align}
& \tilde{p}(\omega,\tau\geq a)
\nn & = 2 \, e^{\frac{i \,\kappa \, a} {2} }
\left[
  \frac{
  e^{\frac{i\, \omega\, (a-\tau)} {2}}
  \sin\!\big(\tfrac{(\omega+\kappa)a} {2} \big)}
  {\omega+\kappa}
  + 
  \frac{
  e^{\frac{i\,\omega\, (a+\tau)} {2}}
  \sin\!\big( \tfrac{(\omega-\kappa)a} {2} \big)}
  {\omega-\kappa}
\right ]
\nn & \quad
+ \frac{2\, e^{\frac{i\,\omega\, a} {2} }}{\omega}
\!\Big[
  \big(e^{i\,\kappa \,a}-1\big)
  \sin \Big( \tfrac{\omega \, (\tau-a)} {2}\Big)
 \nn & \hspace{1.75 cm} -2\cos \Big(\tfrac{\omega \,\tau} {2}\Big)
  \sin \Big(\tfrac{\omega\, a} {2} \Big)
\Big].
\end{align}
 For a meaningful interpretation in terms of single injected excitations, only the regime $\tau>a$ should be considered, as it avoids strong overlap between two incident pulses. In the narrow-pulse limit $a \to 0$, this expression reduces to Eq.~\eqref{eq:psharp}. In contrast to that equation, for finite $a$ the limit $q\to \text{integer}$ yields a finite kernel, allowing a controlled comparison between integer and fractional $q$. We see clearly that the kernel depends on the width $a$, whether $q$ is fractional or integer, leading to a dependence of $S^{HOM}(\tau)$ on $a$. This dependence will be more explicit within the TLL model, in contrast with Refs.~\cite{martin_anyons_dip_HOM_PRL_2023}.

\paragraph{Case of a TLL model.---}  The vast majority of theoretical studies of Hall edge states rely on low-energy effective theories below the UV cutoff $\omega_c$. When a single quasiparticle species dominates, tunneling is governed by one scaling dimension $\delta$, which plays the role of the TLL parameter. In general, $\delta$ may deviate from $\nu$ due to nonuniversal features such as inter-edge interactions or edge reconstruction. For definiteness, we assume quasiparticle tunneling localized at $x=0$. In this case, one expects a quantum metal–insulator transition at an energy scale that separates the weak- and strong-backscattering regimes \cite{kane_fisher_conductance,fendley_prl_1}. The exact relations in Eqs.~\eqref{eq:exact_cross_dc} and \eqref{eq:exact_cross_photo} remain valid across the crossover up to the insulating regime for which $S^{ud}$ vanishes at $\omega_J=0$ and $T=0$.  

Let us now focus on the weak backscattering regime (metallic side), where quasiparticle tunneling is weak. In this case, the dc cross-correlations are given by
\begin{align}
\label{eq:Sdc_ud_integral}
 S_{dc}^{ud}(\omega_J)
 & = 2\,e^*\omega_{th} G_{dc}(0)
 \, \sqrt{\pi} \,\Gamma(\delta+ \frac{1}{2})
\int dx \, \frac{x\sin(2 x\mu )}{\cosh^{2\delta} x},
\nn \text{with }
\mu &= \frac{\omega_J}
{2 \,\pi  \omega_{th}}\,.
\end{align}
Here
\begin{align}
\label{eq:GT}
G_{dc}(0) = \frac{e^{*} \,\mathcal{R}} {2\,\pi} \,
\frac{\Gamma^2(\delta)}{\Gamma(2\delta)}\,(2 \, \pi)^{4(\delta-1) }
\left(\,\frac{\omega_{th}}{\omega_c}\right)^{2 \,(\delta-1)},
\end{align}
is the linear conductance, with $\mathcal{R}\ll 1$, where $\mathcal{R}$ is the reflection probability in the noninteracting limit (i.e., at $\delta=1$). This yields \cite{kyrylo_FQHE_2024}
\begin{align}
\label{eq:Scross_TLL}
S^{ud}_{dc}(\omega_J) &= {2 \,e^* \omega_{th} G_{dc}(0) \, \sqrt{\pi}}
 \nn & \quad
\times \sinh(\pi\mu)\,
\bigl|\Gamma(\delta+i\mu)\bigr|^{2}\,
\Im\psi(\delta+i\mu)\,.
\end{align}

Notice that $S^{ud}_{dc}(\omega)$ decays too slowly for $\delta>1/2$, since at $T=0$ one has $S^{ud}_{dc}(\omega)\propto |\omega/\omega_c|^{2\delta-1}$. Thus, in order to use this TLL expression, the kernel $|\tilde{p}(\tau,\omega)|^2$ in Eq.~\eqref{eq:S_perturbative_HOM} must decay rapidly, requiring $\omega_c\tau \gg 2\pi$. 

Using Eq.~\eqref{eq:S_perturbative_HOM}, we now analyze the HOM noise under sharp pulses. This case is directly motivated by the recent experiment in Ref.~\cite{feve_anyons_martin_Science_2025} where charges $N e/3$ (i.e., $q=N/3$) were injected at $\nu=1/3$ in order to disentangle the scaling dimension $\delta$ from the anyonic braiding phase $\theta$. For $N=1$ ($q=1/3$), the HOM dip width was claimed to scale as $T\delta$, thereby providing a direct probe of $\delta$. For $N=3$, injected electrons do not braid, and the dip width was instead attributed to that of the injected pulses. The contrast between the $N=1$ and $N=3$ cases was then interpreted as a manifestation of anyonic braiding of the injected excitations with thermally created quasiparticle–quasihole pairs at the QPC, from which $\theta=2\pi/3$ was inferred.  This analysis is based on Ref.~\cite{martin_fractional_statistics_dip_HOM_PRL_2023} claiming that while integer-charge pulses yield a dip controlled by their width,  sharp fractional-charge pulses lead to a backscattering noise $S_{ph}(\omega_J=0,\bar{\tau})$ approximated by $1-\exp(-2\delta |\bar{\tau}|)$, where $\bar{\tau}=\pi \omega_{th}\tau $.  We are able to carry out the full analytical calculation of $S_{ph}(\omega_J=0,\bar{\tau})$ via Eq.~\eqref{eq:UNEPT_ph} that reveals significant deviations from this function. It is only when we consider the relevant cross-correlated HOM noise $S^{\mathrm{HOM}}(\tau)$, shown in Fig.~\ref{fig:Fig.2} at $\delta=2/3$, that we more closely approach this functional behavior (see the inset).

\begin{figure}[t!]
\centering
    \includegraphics[width= 0.99 \linewidth]{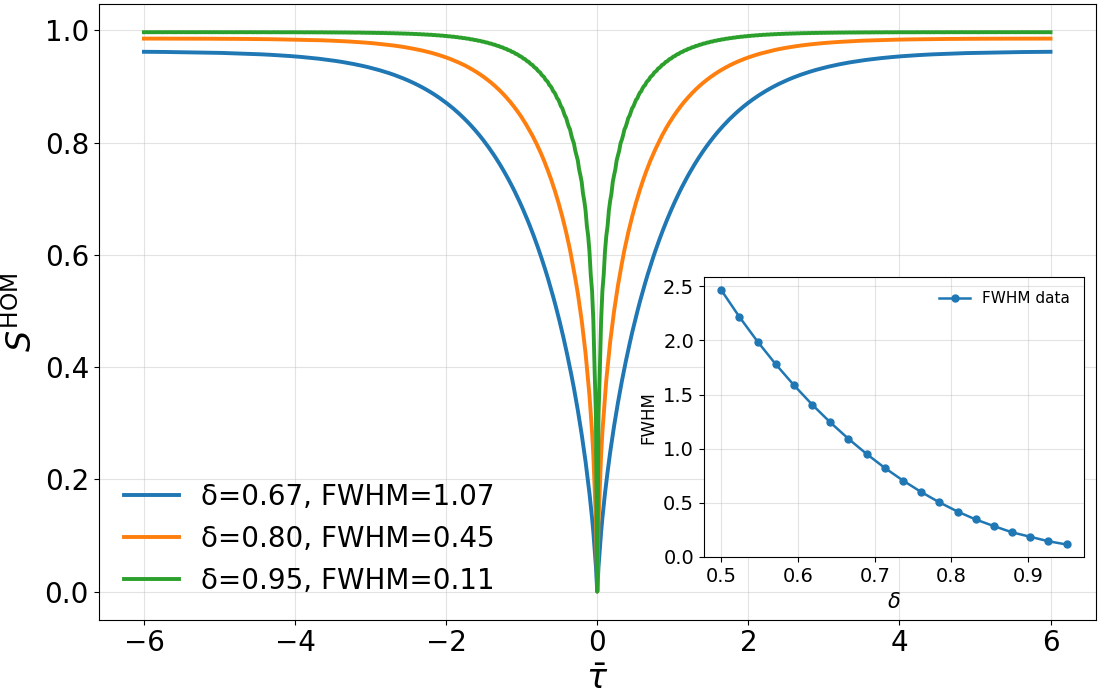}
    \caption{Normalized HOM noise $S^{\mathrm{HOM}}(\bar{\tau})$ as a function of $\bar{\tau}= \pi \omega_{th} \,\tau$ for sharp pulses and for different values of $\delta=\{0.67, 0.80, 0.95\}$. Inset: numerical full width at half maximum (FWHM) of the HOM dip as a function of $\delta$ according to a power law behavior.\label{fig:width_Tdelta}}
\end{figure}

Furthermore, we have extracted the full width at half maximum (FWHM) of the HOM dip, shown in Fig.~\ref{fig:width_Tdelta}. 
The data reveal that the dip width is predominantly governed by the scaling dimension $\delta$. 
More precisely, it decreases markedly with increasing $\delta$, 
demonstrating a pronounced power-law behavior (see inset). 
This establishes a robust and universal scaling of the HOM feature with respect to noninteger charges $q$.

\begin{figure}[t!]
    \centering
\includegraphics[width= 0.99 \linewidth]{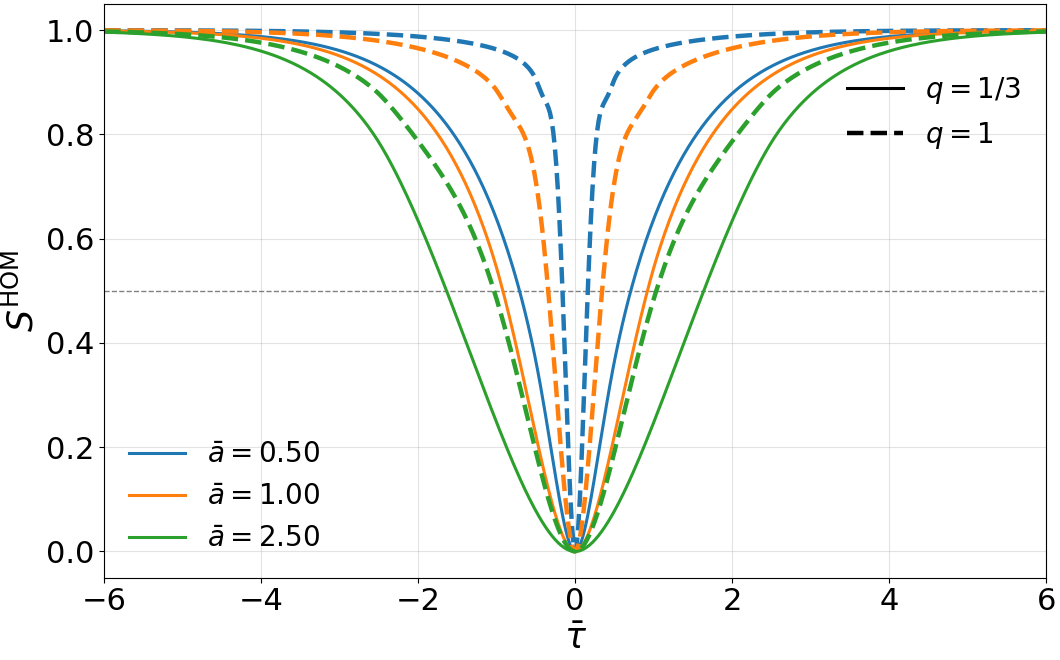}
    \caption{Normalized HOM noise $S^{\mathrm{HOM}}(\bar{\tau})$ as a function of the delay $\bar{\tau}$ for rectangular pulses of finite width $\bar{a}=\pi\omega_{th} a$, for integer $q=1$ (dashed) and fractional $q=1/3$ (solid), respectively. \label{fig:Fig.4}}
\end{figure}

By analogy with the sharp-pulse case, we compute $S^{\mathrm{HOM}}(\tau)$ for rectangular pulses of various widths $\bar{a} = \pi \omega_{th} a$, as illustrated in Fig.~\ref{fig:Fig.4}. As expected, the HOM signal exhibits a strong dependence on $\bar{a}$, and the limit $\bar{a}\to0$ formally reproduces the sharp-pulse result shown in Fig.~\ref{fig:Fig.2}. This dependence holds for both fractional and integer excitations, in contrast to Refs.~\cite{martin_anyons_dip_HOM_PRL_2023,feve_anyons_martin_Science_2025}.

\begin{figure}[t!]
    \centering
\includegraphics[width= 0.99 \linewidth]{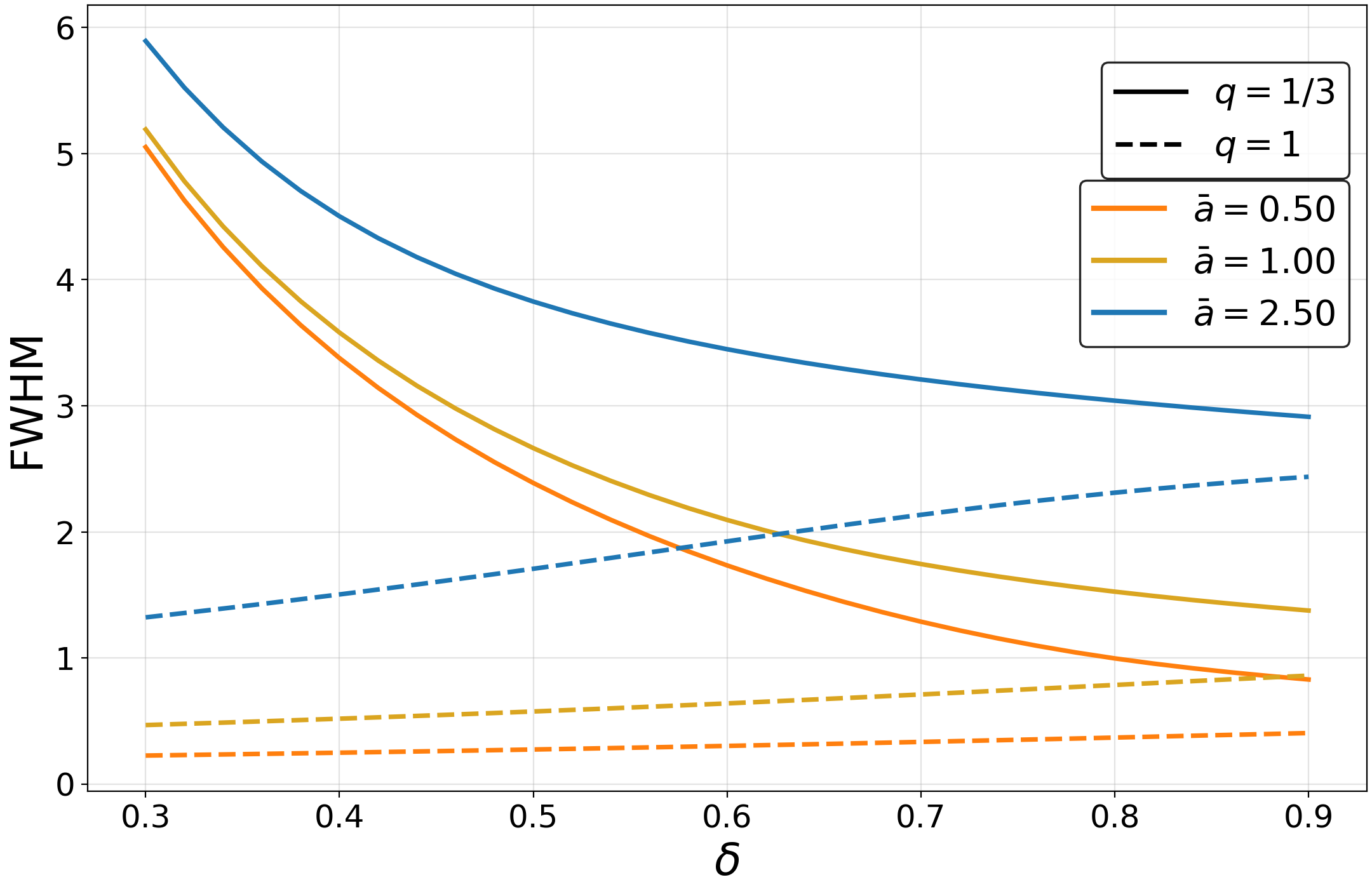}
      \caption{Full width at half maximum (FWHM) of the HOM dip versus scaling dimension $\delta$ for fractional $q=1/3$ (solid line) and integer $q=1$ (dashed line) at $\omega_{th}/\omega_c=0.02$ and $0.1$. \label{fig:Fig.5new}}
\end{figure}

Now we address the dependence of FWHM on the scaling dimension $\delta$ at three different values of the injected pulse widths $a$. Figure~\ref{fig:Fig.5new} summarizes the results for several pulse widths $\bar a=\pi \omega_{th} a$ and both for fractional ($q=1/3$) and integer ($q=1$) charge injection. 
For fractional excitations, the FWHM decreases monotonically with $\delta$ over the entire range explored. Narrow pulses ($\bar a=0.5$) exhibit the steepest decay, whereas broad pulses ($\bar a=2.5$) lead to a smoother, less $\delta$-sensitive profile. 
This behavior reflects the dominance of high-frequency components in the correlation spectrum $S_{\mathrm{dc}}^{\mathrm{ud}}(\omega_{J}=\omega)$ at larger $\delta$, which enhance temporal decoherence and compress the interference feature in delay $\tau$. 
For integer excitations,  the FWHM is nearly constant for small $\bar a$, showing only a weak residual increase with $\delta$ at larger pulse widths. 
Overall, the comparison between $q=1/3$ and $q=1$ highlights a clear dichotomy: fractional excitations lead to pronounced scaling of the HOM width with $\delta$, while integer excitations yield a nearly flat baseline in the short-pulse regime. 
Increasing $\bar a$ broadens the dip but also reduces the contrast between the two charge sectors, demonstrating how pulse shaping and quasiparticle charge jointly govern the temporal width of HOM interference.

\paragraph{{Conclusion.---}} We have introduced a generalized theoretical framework for photo-assisted collisional interferometry in fractional quantum Hall edge states. By extending non-equilibrium bosonized edge theory (NEBET) to time-dependent sources, we derived exact relations connecting cross-correlated photo-assisted noise to backscattering noise, independent of microscopic details of the bosonized Hamiltonian or tunneling operator. A key universal feature is that HOM noise vanishes at zero delay.  

Specializing to the chiral TLL model, we analyzed the HOM dip produced by pairs of sharp pulses, showing that the normalized HOM signal is independent of noninteger injected charge $q$.  Extending the analysis to rectangular pulses of finite width, we derived explicit analytic expressions for the drive kernel and showed that the pulse width introduces an additional scale that broadens the HOM dip and gradually reduces its sensitivity to $\delta$, smoothing the contrast between fractional and integer excitations. Altogether, these results clarify how interaction strength, quasiparticle charge, and pulse shaping interplay to determine HOM interference features, offering robust predictions for future experiments aimed at probing anyonic statistics in strongly correlated systems. 
\paragraph{Acknowledgments:} I. S. thanks Lucas Mazzella,  Seddik Ouacel, Gu Wang, Igor Gornyi, Gabriele Campagnano, Pascal Degiovanni and Gwendal F\`eve for discussions and ongoing collaboration.

\bibliographystyle{apsrev4-2}

%

\end{document}